\documentclass{aa}   
\usepackage{graphicx} 
\begin{document} 
\title{The Clustering Evolution of Distant Red Galaxies in the 
GOODS-MUSIC Sample} 
 
   \author{A. Grazian 
          \inst{1} 
          \and 
          A. Fontana\inst{1} 
          \and 
          L. Moscardini\inst{2} 
          \and 
          S. Salimbeni\inst{1} 
          \and 
          N. Menci\inst{1} 
          \and 
          E. Giallongo\inst{1} 
          \and 
          C. De Santis\inst{1} 
          \and 
          S. Gallozzi\inst{1} 
          \and 
          M. Nonino\inst{3} 
          \and 
          S. Cristiani\inst{3} 
          \and 
          E. Vanzella\inst{3} 
          } 
 
   \offprints{A. Grazian, \email{grazian@oa-roma.inaf.it}} 
 
\institute{INAF - Osservatorio Astronomico di Roma, Via Frascati 33, 
I--00040 Monteporzio (RM), Italy 
\and Dipartimento di Astronomia - Universit\`a di Bologna, Via Ranzani 1, 
I--40127 Bologna, Italy 
\and INAF - Osservatorio Astronomico di Trieste, Via G.B. Tiepolo 11, 
I--34131 Trieste, Italy}

   \date{Received .... ; accepted ....} 
   \titlerunning{Clustering of DRGs in the GOODS-South} 
  
  \abstract 
  % context heading (optional) leave it empty if necessary   
{ 
} 
{We study the clustering properties of Distant Red Galaxies (DRG) 
to test whether they are the progenitors of local massive  
galaxies. 
} 
{We use the GOODS-MUSIC sample, a catalog of 
$\sim$3000 Ks-selected galaxies based on VLT and HST observation of the
GOODS-South field with extended multi-wavelength 
coverage (from $0.3$ to $8 \mu m$) and accurate estimates of the 
photometric redshifts to select 179 DRGs with 
$J-Ks\ge 1.3$  in an area of 135 sq. arcmin.} 
{We first show that the $J-Ks\ge 1.3$ criterion selects a rather 
heterogeneous sample of galaxies, going from the targeted high-redshift
luminous evolved systems, to a significant 
fraction of lower redshift ($1<z<2$) and less luminous dusty starbursts. 
These low-redshift DRGs are significantly less clustered than higher-$z$ 
DRGs. With the aid of  extreme and simplified theoretical models of
clustering evolution we show 
that it is unlikely that the two samples are drawn from the same 
population observed at two different stages of evolution.  } 
{High-$z$ DRGs likely represent 
the progenitors of the more massive and more luminous galaxies in the 
local Universe and might mark the regions that will later evolve into 
structures of intermediate mass, like groups or small galaxy clusters. 
Low-$z$ DRGs, on the other hand, will likely evolve into slightly less 
massive field galaxies.} 
 
   \keywords{Galaxies: distances and redshift - Galaxies: evolution - 
Galaxies: high redshift} 
 
   \maketitle 
% 
%________________________________________________________________ 
 
\section{Introduction} 
 
Finding and studying large samples of distant luminous and evolved 
galaxies is fundamental to provide a deeper insight on the formation 
of massive galaxies, a process that is commonly perceived as a 
challenging test for cosmological models of structure formation and 
evolution. For this reason, in the recent past, the study of early 
type galaxies at the highest observable redshifts made use of a 
considerable fraction of large telescope time and occupied a 
substantial part of the astronomical literature. 
 
The search for passively evolving systems at high redshift began with 
the so-called Extremely Red Objects (EROs; see also 
\cite{rieke,cimatti,mcarthy}) which reproduce the colours of 
ellipticals at $z\sim 1$.  EROs are relatively ``new'' objects, in the 
sense that they have been recognized as a specific class only around 
1990, due to the availability of Near InfraRed (NIR) detectors only at 
that epoch.  Elston, Rieke \& Rieke (1988) found the first two EROs in 
a survey of 10 sq. arcmin. as resolved galaxies with $K\sim 16.5$ and 
$R-K\sim 5$.  After the optical spectroscopic identifications, the two 
objects turned out to be an evolved galaxy at $z=0.8$ and a dusty 
starburst at $z=1.44$, named HR10. It was clear from this survey and 
successive investigations that the ERO population is heterogeneous in 
its main properties (star formation, mass, age, extinction, etc.). 
 
At present, there are various techniques to find evolved galaxies at 
high-$z$. \cite{cimatti99} utilised the criterion $R-K\ge5$(Vega) 
effective in the redshift interval $0.8\le z\le 1.8$ limited to $Ks\le 
20$.  \cite{caputi} and \cite{abraham} used a similar selection 
$I-K\ge4$(Vega) to select red galaxies with $1\le z\le 2$. \cite{pm00} 
suggested a two colour criterion ($I-K$ vs $J-K$) to separate ellipticals 
from dusty starbursts at $1\le z\le 2$, which could be extended at 
higher redshift ($2.0\le z\le 2.5$) using redder colours ($J-K$ vs 
$H-K$). \cite{franx} proposed a simple pure infrared criterion 
$J-K\ge2.3$(Vega) for $z\ge 2.0$. In a similar way, \cite{saracco04} 
selected 3 galaxies with $J-Ks\ge 3$(Vega) in the HDFS at $z\ge 2.5$, 
plausible candidates for high-z massive galaxies, though the statistic 
is very limited. Recently, \cite{bzk} suggested to 
isolate early-type galaxies according to the BzK criterion 
[$(z-K)_{AB}-(B-z)_{AB}\le -0.2$ and $(z-K)_{AB}\ge 2.5$] efficient at 
$1.4\le z\le 2.5$, and with extension at $2.5\le z\le 4.0$ using the 
RJL colour combination. \cite{yan} proposed a new class of objects, the 
high-$z$ EROS (called IEROs) with $f_\nu(3.6\mu)/f_\nu(850nm)\ge 20$ 
to select red galaxies at $1.5\le z\le 3.0$ using MIR data. 
The physical properties of massive galaxies at high-$z$ were also 
investigated by \cite{saracco05} through spectroscopy of a limited sample of 
massive, evolved galaxies with relatively bright magnitudes ($K\le 18.4$) at 
$1.3\le z\le 1.7$ on the MUNICS survey. 
A different approach has been adopted for the COMBO-17 survey, in which 
the intrinsic colour (U-V) rest frame is utilised to isolate galaxies 
belonging to the Red Sequence: \cite{combo} used the relation 
$(U-V)_{\rm rest}\ge 1.40-0.31\cdot z$, efficient at $0.2\le z\le 1.1$ 
according to simulations with spectral synthesis code. Finally, 
\cite{giallongo} adopted a slightly different approach: the 
bi-modality in $(U-V)_{\rm rest}$ is empirically fitted to the 
observations and could be extended up to $z\sim 3$. 
 
In this paper we focus on the so-called Distant Red Galaxies (DRGs; 
\cite{franx}).  These galaxies are selected through a $J-K>2.3$(Vega) 
criteria, designed to be sensitive to galaxies with a large 4000 \AA~ 
break at $z\ge 2$. \cite{franx} used this technique in the FIRES 
survey (\cite{labbe03}) selecting 14 DRGs in the HDFS, down to faint 
$Ks$ magnitudes ($Ks\le 24.5$ in AB mag).  By using ultra-deep 
spectroscopy \cite{vandokkum} provided evidence that the brighter DRGs 
are indeed galaxies at $z\sim 2$. Even if the evidence for the 
existence of old and massive galaxies is settled by these 
observations, however the lack of a statistical significant sample of 
DRGs hampered the detailed study of many of their properties, in 
particular their number density, their clustering properties and their 
physical properties like mass, star formation, age and spectral 
energy distribution (SED). Recently, 
\cite{papovich05} have derived a sample of 153 DRGs from the GOODS 
South down to a shallower limit of $Ks=23.2(AB)$, with the aim of 
studying in detail the specific star formation rate (star formation 
per unit mass star) of DRGs. They found that the bulk of the star 
formation in massive galaxies ($M\ge 10^{11} M_{\odot}$) occurs at 
early cosmic epochs and is largely complete by $z\sim 1.5$. 
 
Analogously to \cite{papovich05}, we use the extraordinary dataset 
provided by the GOODS survey to extend these studies. In particular, 
we will adopt the GOODS-MUSIC sample, a $Ks$-selected sample with an 
extended wavelength range (from U to $8.0\mu$m band) that we compiled 
using publicly available data in the Chandra Deep Field South region 
and described at length in \cite{grazian}. With this complete 
sample of DRGs, we can define in detail their general properties and 
refine previous investigations by \cite{franx}, which used only 14 
objects in the FIRES survey, though at a fainter magnitude limit. 
 
The structure of the paper is as follows. In \S\ 2 we describe the 
data used to analyse DRG properties. In \S\ 3 we select DRGs according 
to the selection criterion defined by \cite{franx} and provide their 
number density and the redshift distribution. In \S\ 4 we study the 
clustering properties of DRGs selected in the GOODS-South field and 
in \S\ 5 we discuss the link between the DRG population at $z\sim 2$ 
and the local ellipticals. 
 
All magnitudes, unless otherwise stated, are given in the AB system. 
A concordance $\Lambda CDM$ cosmological model ($\Omega_M =0.3$, 
$\Omega_\Lambda =0.7$ and $H_0 = 70$ km s$^{-1}$Mpc$^{-1}$) has been 
adopted throughout the paper. 
 
%__________________________________________________________________ 
 
\section{The Data} 
 
We use in this paper the data from the Chandra Deep Field South (CDFS; 
\cite{giacconi}), obtained within the GOODS survey. This is a 
collaboration between STScI and ESO (\cite{renzini}) that produced an 
unprecedented dataset of images, covering 135 sq. arcmin. from 0.3 to 
8.0$\mu$m down to relatively faint magnitude limits 
(\cite{giavalisco}).  In particular, we used the ACS images (release 
V1.0, \cite{giavalisco}), the ISAAC database (release V1.0, 
\cite{vandame}) and the IRAC dataset (release V1.0 enhanced, 
\cite{dickinsonirac}), together with U band photometry from WFI@2.2m 
ESO-MPI and VIMOS reduced by our group. 
 
Using this public dataset, we have produced a high-quality multicolour 
catalog of galaxies in the GOODS-South, that we have named 
GOODS-MUSIC: details about the procedure adopted are discussed in 
\cite{grazian}. We briefly remind here that we have used all the 
publicly available images from U to 8.0 $\mu$m 
($U,B,V,i,z,J,H,Ks,3.6\mu,4.5\mu,5.8\mu,8.0\mu$), in a contiguous area 
of 135 sq. arcmin., totalling 14847 objects.  In 
particular, to isolate a complete sample of DRGs, we use here the 
$Ks$-selected sample, that consists of $2931$ galaxies. 
The GOODS survey has a complex, inhomogeneous exposure map in the $Ks$ 
band. To properly derive the statistical properties of galaxies in 
this field, the sample has been divided in 6 sub-areas of different 
magnitude limits, as described in details in \cite{grazian} and in 
Tab. \ref{khisttab}. This information is used in this work when the 
DRG statistical properties are studied, such as their number density 
or clustering properties. The typical magnitude limit for most of the sample
is about $Ks=23.5$, and extends down to 23.8 in a limited area.

\begin{figure} 
\includegraphics[width=9cm]{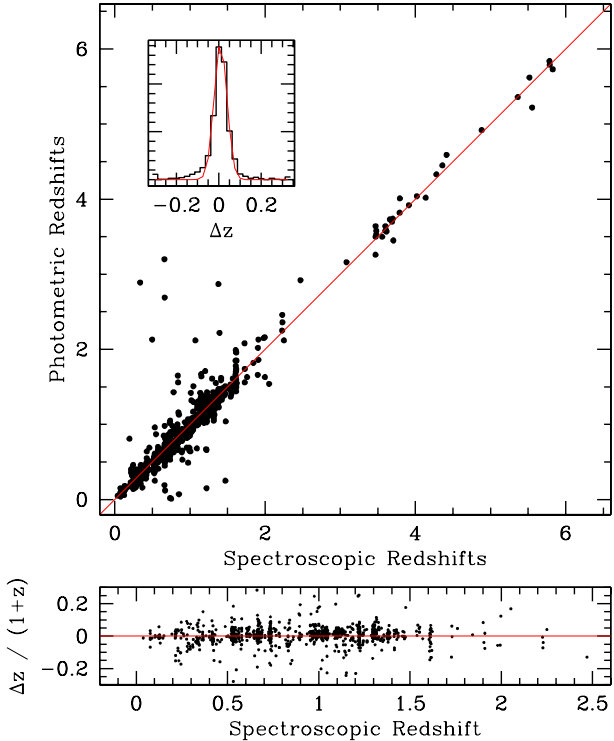} 
\caption{The spectroscopic vs photometric redshifts for 973 galaxies 
in the GOODS-MUSIC sample. The accuracy is $\sigma_z=0.06$ and 
$\frac{\sigma_z}{1+z}=0.03$ in the redshift range $0<z<6$. 
} 
\label{zszp} 
\end{figure}

\begin{figure} 
\includegraphics[width=9cm]{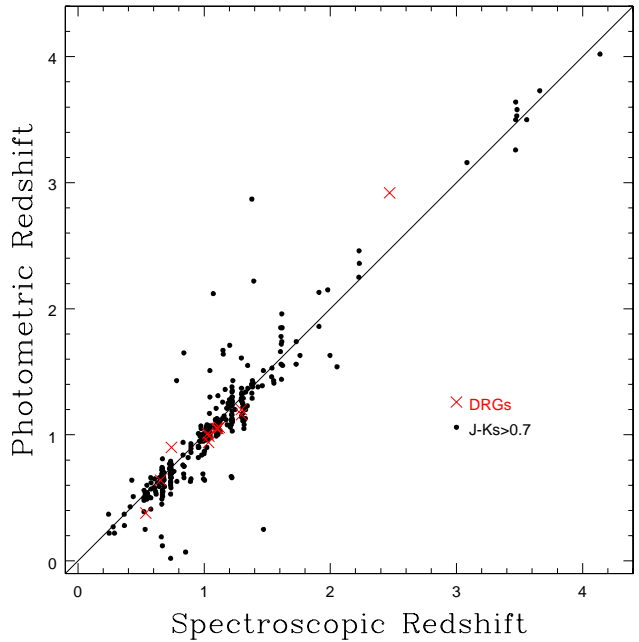} 
\caption{The spectroscopic vs photometric redshifts for 340 red galaxies 
with $J-Ks\ge 0.7$ in the GOODS-MUSIC sample. The accuracy is 
$\sigma_z=0.08$ and $\frac{\sigma_z}{1+z}=0.05$ in the redshift 
range $0<z<4$. There are only 13 galaxies with $J-Ks\ge 1.3$ and 
spectroscopic redshifts (red crosses). 
} 
\label{drgzszp} 
\end{figure}

In \cite{grazian} we included spectroscopic information for 668 
galaxies. Recently, \cite{vanzrun2} have released further 
spectroscopic redshifts in the GOODS South region. We used this new 
release to compile a revised sample of 973 galaxies with good 
spectroscopic identification.
Out of this number, 815 are in the $Ks$-selected sample
($\simeq 28 \%$ of the total). For the remaining 
sources, we derived a photometric redshift, as described in 
\cite{grazian}: the redshift accuracy in the range $0<z<6$, as shown in 
Fig.\ref{zszp}, on this enlarged spectroscopic 
sample is $\sigma_z=0.06$, which is the same
value previously found in \cite{grazian}. If 
we restrict to the 340 galaxies with red colours ($J-Ks\ge 0.7$), as 
shown in Fig.\ref{drgzszp}, the redshift accuracy is $\sigma_z=0.08$ 
in the redshift range $0<z<4$.

Rest--frame physical quantities (such as luminosities, mass, age, SFR) 
are derived by using the synthetic library of \cite{bc03} (hereafter 
BC03), at the spectroscopic redshift, adopting the same technique 
already described in several previous papers (see Fontana et al. 2004 
for more details). 
 
%__________________________________________________________________ 
 
\section{Selection of DRGs} 
 
\subsection{The number density of DRGs} 
 
\begin{figure}
\includegraphics[width=9cm]{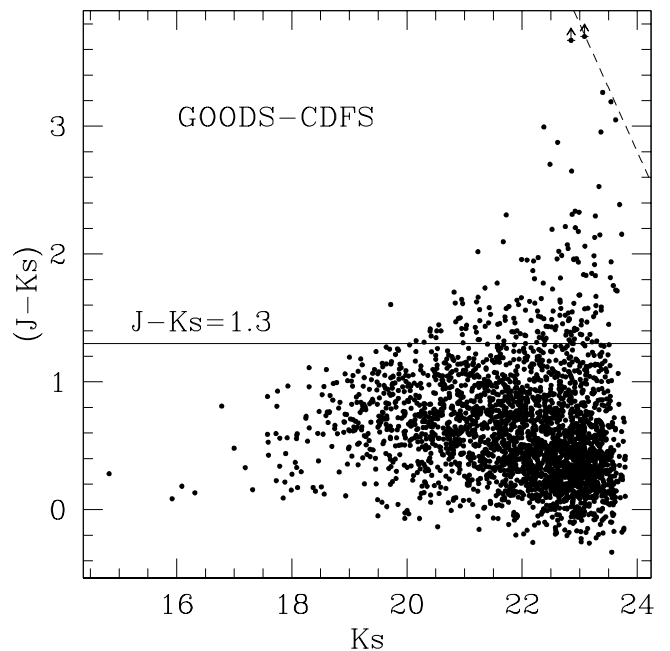} 
\caption{Selection of $J-Ks\ge 1.3$ objects in the GOODS-South Field. 
%The sample is complete down to $Ks=23.8(AB)$. 
Upper limits in the $J$ band are shown as vertical arrows. The 
horizontal line shows the selection criteria for DRGs in the GOODS-South 
area, while dashed line indicates the completeness on the DRG selection 
due to the depth of the J band (26.8 AB at $S/N=1$). 
} 
\label{JKvK} 
\end{figure}

We have selected DRGs according to the criterion defined by 
\cite{franx}, ($J-Ks\ge 1.3$ in AB system, as obtained using the 
transmission curves for the J and Ks filters of ISAAC), which is 
efficient at $z\ge 2$.  Fig. \ref{JKvK} shows the effect of this 
selection criterion of DRGs applied to the objects of GOODS-MUSIC 
sample. 
 
In the GOODS-South region we find 179 galaxies having $J-Ks\ge 1.3$.
For the reasons described above, the completeness limit of the
survey is not homogeneous, with a typical value of $Ks=23.5$. We 
use this sample of DRGs to study in particular their number density and 
their spatial distribution (clustering). 
 
The number density of DRGs in the GOODS South field is derived through 
the classical $\log N-\log S$ distribution, or the number of objects 
per sq. arcmin.  and per magnitude bin in the $Ks$ band. This last 
quantity is obtained by following the recipe of \cite{avni}: 
 
\begin{equation} 
n(Ks)=\frac{1}{\Delta Ks}\sum_{i=1}^{N_{\rm obj}} \left[ \sum_{j=1}^{N_{\rm  
field}} 
Area_j^{\rm max} \right]^{-1} \ , 
\end{equation} 
where the sum is on the $N_{\rm field}$ surveys (here, the 6 areas with 
different magnitude limits described in \cite{grazian} and in 
Tab.\ref{khisttab})
and on the $N_{\rm obj}$ objects; $Area_j^{\rm max}$ represents the 
accessible area of 
the $j$-th survey (this is equivalent to the maximum accessible volume 
when the luminosity function is derived). The DRG counts have been 
computed in bins of $\Delta Ks=0.5$ magnitude.
 
\begin{figure} 
\includegraphics[width=9cm]{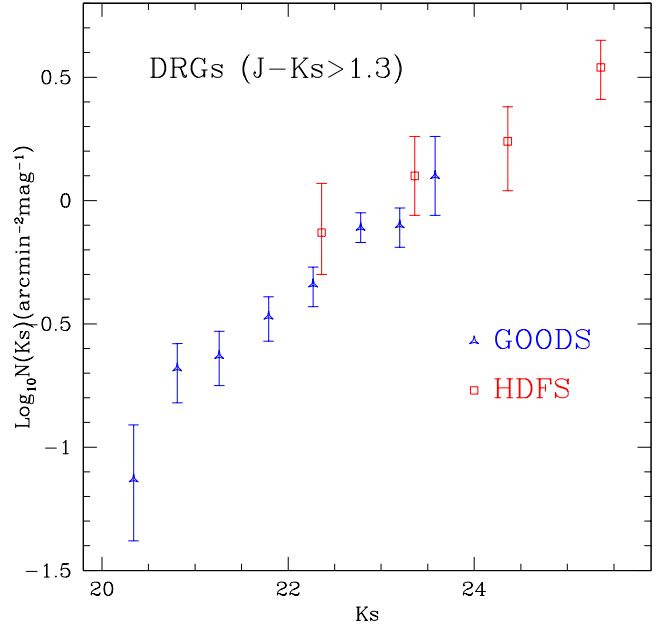} 
\caption{The surface density of selected DRGs in the GOODS-South Field 
(triangles), compared with the estimate obtained in the HDFS (squares, 
\cite{labbe03}).}
\label{logns} 
\end{figure}

Fig. \ref{logns} shows the surface density of DRGs in the GOODS South 
field and compares it with the results derived in the HDFS by the 
FIRES survey (\cite{labbe03}). Even if the area of HDFS is smaller 
with respect to the GOODS Survey, the DRG number densities in these 
two independent fields are comparable (see also Table \ref{khisttab}). 
Notice, however, that different values for the number density of DRGs 
has been derived by using the data of the HDFN 
(\cite{lanzetta,fontana,dickinson}), in which one DRG is found at $Ks\le 
23.0$, and a limited number at $23\le Ks\le 24$ with upper limit in 
the $J$ band. The sample variance between HDFN and HDFS is due to the 
limited area investigated and stresses the necessity of deriving a 
firm measurement for the number density of DRGs in a large and deep 
survey such as the GOODS-CDFS field. 
 
\begin{table} 
\caption[]{Number density $\Sigma$  
of DRGs in the $Ks$ band for GOODS-South and HDFS 
fields} 
\begin{tabular}{lccccc} 
\hline 
\hline 
$bin$ & $N$ & $\log(\Sigma)$ & $\bar{z}$ & $\bar{Ks}$ & {\small AREA} \\ 
 &   & mean $+1\sigma$ $-1\sigma$ & & & $arcmin^2$ \\ 
\hline 
20.25 &  6 & -1.13 -0.91 -1.38 & 1.05 & 20.34 & 135.372 \\ 
20.75 & 14 & -0.68 -0.58 -0.82 & 1.25 & 20.81 & 135.372 \\ 
21.25 & 16 & -0.63 -0.53 -0.75 & 1.42 & 21.26 & 135.372 \\ 
21.75 & 22 & -0.47 -0.39 -0.57 & 1.96 & 21.79 & 129.692 \\ 
22.25 & 29 & -0.34 -0.27 -0.43 & 2.04 & 22.27 & 128.273 \\ 
22.75 & 50 & -0.11 -0.05 -0.17 & 2.45 & 22.78 & 127.935 \\ 
23.25 & 32 & -0.10 -0.03 -0.19 & 2.80 & 23.20 &  81.272 \\ 
23.75 & 10 &  0.10  0.26 -0.06 & 2.75 & 23.58 &  12.585 \\ 
\hline 
22.50 &  4 & -0.13  0.07 -0.30 & 2.75 & 22.36 &   4.500 \\ 
23.50 &  7 &  0.10  0.26 -0.06 & 2.75 & 23.36 &   4.500 \\ 
24.50 &  8 &  0.24  0.38  0.04 & 2.75 & 24.36 &   4.500 \\ 
25.50 & 18 &  0.54  0.65  0.41 & 2.75 & 25.36 &   4.500 \\ 
\hline 
\end{tabular} 
\label{khisttab} 
\begin{list}{}{} 
\item a) 
the number density $\Sigma$ is in units of $arcmin^{-2} mag^{-1}$ 
\item b) 
$bin$ represents the central bin  magnitude in $Ks$ 
\item c) 
$\bar{z}$ and $\bar{Ks}$ are the mean values of redshift and observed $Ks$ 
magnitude for each magnitude bin 
\item d) 
the number density in the second half of the table derives 
from the FIRES survey in the HDFS (\cite{labbe03}) 
\end{list} 
\end{table} 
 
\subsection{Redshift distribution of DRGs} 
 
\begin{figure}
\includegraphics[width=9cm]{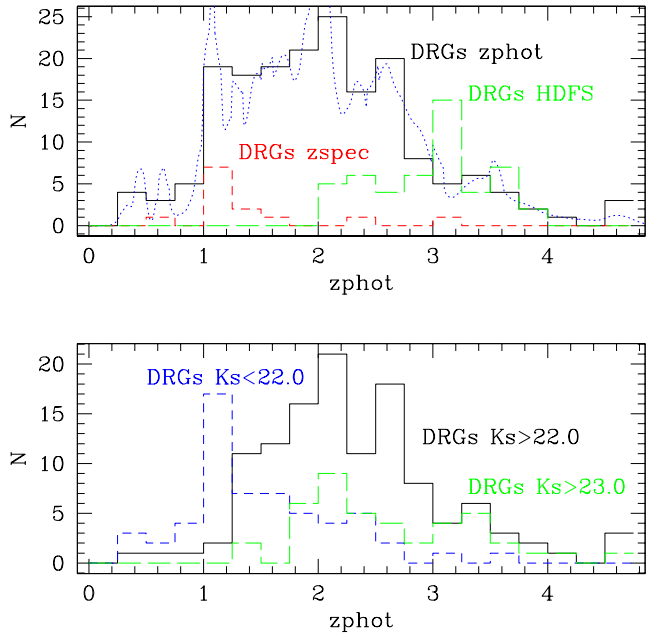} 
\caption{{\em Upper panel}: the distribution  
of spectroscopic (only 13 objects; short-dashed line) and photometric (solid 
line) redshifts of selected DRGs in the GOODS-South Field. The dotted curve 
is the redshift distribution obtained for the DRGs using the probability 
function for the redshift for each object derived by the photometric 
redshift code. It is in agreement with the distribution using the best
estimate 
for the photometric redshift code. The long-dashed line represents the 
redshift distribution for the HDFS (\cite{labbe03}), peaked at $z\sim 3$. 
It is markedly different from the redshift distribution of the GOODS field, 
since DRGs in the HDFS have fainter $Ks$ magnitudes. The redshift 
distribution of HDFS is comparable to the redshift distribution of DRGs in the 
GOODS field at $Ks>23$ magnitude, as shown in the lower panel. 
{\em Lower panel}: the photometric redshift distribution for bright ($Ks<22$; 
long-dashed line) and faint ($Ks>22$; solid line) DRGs. Deep pencil beam 
surveys (HDFs) preferentially select objects at $z\sim 2$, while large 
area surveys are biased towards lower-redshift ($z\le 2$) and bright
($Ks<22$) DRGs (short-dashed line).} 
\label{zhist} 
\end{figure} 
 
\begin{figure} 
\includegraphics[width=9cm]{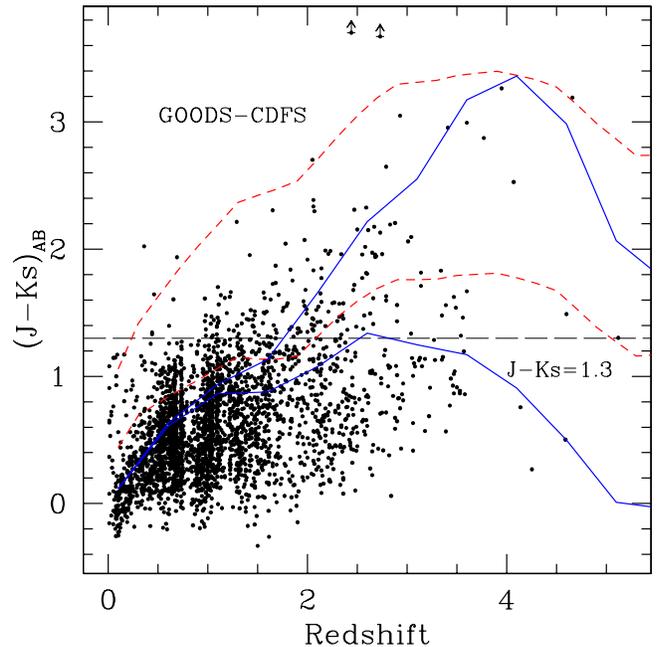} 
\caption{The $J-Ks$ colour of objects in the GOODS-South field as a 
function of their (spectroscopic or photometric) redshift.  Upper 
limits in the $J$ band are displayed as vertical arrows.  The 
long-dashed horizontal line shows the selection criteria adopted for 
DRGs in this paper. The two blue solid lines show the $J-Ks$ colour for 
passively evolving galaxies formed at $z=20$ and with an e-folding 
star formation rate with timescale $\tau=0.1$ and $\tau=1$ Gyr (upper 
and lower curves, respectively). The red short-dashed lines show the 
same colour for a star-forming galaxy with $E(B-V)=1.1$ and 
$E(B-V)=0.5$ (upper and lower curves, respectively).  } 
\label{JKvred} 
\end{figure} 
 
The large number of DRGs in the GOODS field makes it possible to test 
the selection criterion and to define the window function in redshift 
for DRGs. Fig. \ref{zhist} shows the distribution of the 
photometric redshifts of DRGs: the spectroscopic sample of DRGs is 
very limited both in redshifts and $Ks$ magnitudes (only 13 galaxies 
with $19.7\le Ks\le 22.9$ and $ 0.65\le z\le 3.04$). The redshift 
distribution of GOODS DRGs is slightly different from that drawn for 
HDFS by \cite{franx} and \cite{daddi}, which covers the interval $2\le z\le 4$ 
with a prominent peak at $z\sim 3$, and in reasonable agreement with 
the similar analysis of Papovich et al 2005. In our GOODS-MUSIC sample 
there are DRGs at lower redshifts ($1\le z\le 2$) with bright apparent 
$Ks$ magnitudes ($Ks\le 22$), which are in practice absent in small 
and deep pencil beam surveys, like the HDFS.  The redshift 
distribution clearly shows that there is a considerable fraction (77 
out of 179, i.e. 43$\%$) of objects at low redshifts ($z\le 1.9$) 
which satisfy the $J-Ks$ selection. With a typical colour $J-Ks\sim 
1.5$, they cannot be the result of photometric errors, since this 
should be negligible for relatively bright objects: in fact at $Ks\sim 
21.5$ the typical error in magnitude is $\sigma=0.03$.  The SEDs of 
these low-redshift DRGs are dominated by power-law spectra with a tilt 
at $\lambda\sim 6\mu m$, which are mostly fitted by relatively young 
galaxies (${\rm age}/\tau\le 1$) and a substantial amount of 
extinction ($E_{B-V}\sim 0.5-1.0$, see Fig. 8 of \cite{papovich05}). 
 
Fig. \ref{JKvred} may help in understanding this result, which is due 
to the complex selection effects that are effective in this colour 
criterion. In Fig. \ref{JKvred} we compare the observed $J-Ks$ colour as 
a function of redshift with the expected $J-Ks$ of a few, selected 
templates computed with the BC03 models.  Two of these models are 
computed adopting exponentially declining star-formation histories, 
both started at very high redshift ($z_{\rm form}=20$), with solar 
metallicity. The values adopted for the e-folding timescale ($\tau=0.1$ 
and $\tau=1$ Gyr) both produce the same colour at low redshift and 
show that the $J-Ks>1.3$ threshold is effective in selecting galaxies 
at $z>2$ that formed their stars in a short starburst $\tau\le 1$Gyr. 
At the same time, large $J-Ks$ colours may be obtained  
by star-forming, dusty models down to lower redshift $z\simeq 1$. 
 
This highlights why the DRG population is not a unique class of $z>2$ 
objects, but it is contaminated by dusty starbursts with $z\sim 1.5$, 
whose strong dust absorption is responsible for their red infrared 
colours. The low-redshift DRG sub-sample is at the limit of the $J-Ks$ 
selection, and can be explained by dust reddening of $z\sim 1.5$ 
star-forming galaxies, as shown in Fig. \ref{JKvred}.  If a more 
drastic cut $J-Ks$ colour would be applied (e.g. $J-Ks\ge 1.8$), this 
would ensure a much more efficient selection of galaxies with $z\ge 
2$, but the sample would be strongly reduced, from 179 to 51 galaxies 
only. 
 
\begin{figure} 
\includegraphics[width=9cm]{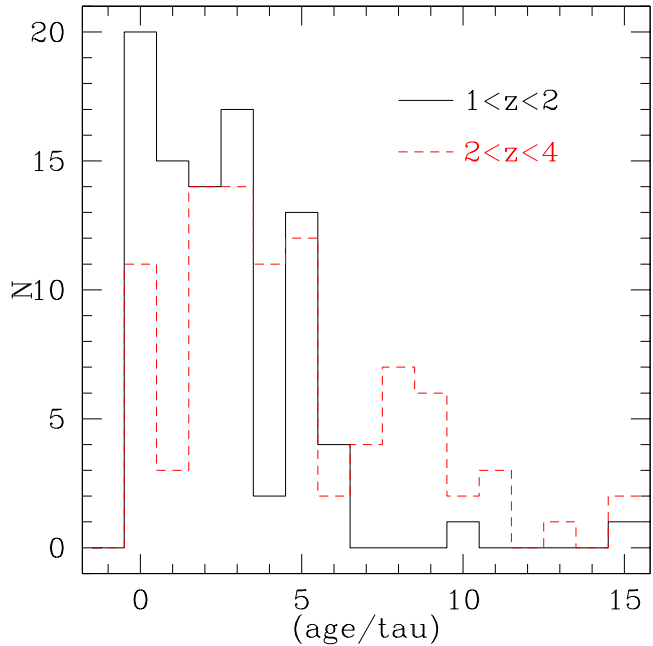} 
\caption{The distribution of the ratio between the age of DRGs and 
the characteristic timescale $\tau$ of the exponentially declining 
SFR, according to the BC03 spectral synthesis model. The solid 
histogram refers to the distribution of low-$z$ DRGs, dominated by 
relatively young objects (age/$\tau\le 3$) which are typically dusty 
starbursts, while the dashed histogram shows the ratio for $2\le z\le 
4$ DRGs, where a considerable fraction (30\%) of old and passively 
evolving galaxies arise.} 
\label{ttau} 
\end{figure} 
 
The difference between low-$z$ and high-$z$ DRG has been extensively 
discussed in a recent paper by \cite{papovich05}. They argue that lower-$z$ 
DRG are dominated by dusty starbursts, while the higher-$z$ 
objects are made of a more complex stellar population, likely a 
mixture of star-forming, heavily extincted and older, passively evolving 
stellar components, with a minority of galaxies that are likely 
genuinely passively evolving. In our preliminary, simplified analysis 
(the most important difference with respect to \cite{papovich05} is that we 
do not use models with two-component stellar populations and we do 
not include the $24 \mu$ data in the analysis) we also have evidence 
of the same distinction. This is shown in Fig.\ref{ttau}, where we 
report the distribution of the ratio between the fitted age and the 
fitted star-formation e-folding timescale $\tau$ (such a ratio is in 
practice the inverse of the Scalo parameter). As it is shown, all low-$z$ 
DRG are dominated by actively star-forming, relatively young 
objects, while higher-$z$ DRGs have a broader distribution of 
age$/\tau$, including several objects (30\% of the high-$z$ DRG sample) 
that are fitted by passively evolving models. 
 
The average luminosities in the rest--frame I band (Vega system) that 
we infer from the spectral fitting of our sample are $<M_I>=-22.3$ and 
$<M_I>=-23.2$ at $<z>=1.5$ and $<z>=2.7$, respectively, and the 
average stellar masses are $<M_*>=8.15\cdot 10^{10} M_\odot$ and 
$<M_*>=9.90\cdot 10^{10} M_\odot$ (10.76 and 10.88 if we compute 
$<\log(M)>$), respectively. 
 
It is tempting to speculate on the possible spectral evolution of 
these objects. A lower limit to their local luminosity can be obtained 
by assuming that they enter into a passive evolution phase soon after 
we observe them. In this case, assuming a truncated star-formation 
history with solar metallicity, the BC03 code predicts in the 
rest-frame I band a fading from $<z>=1.5$ and $<z>=2.7$ to $z\simeq 0$ 
of 2.2 and 2.45 magnitudes, respectively.  However, we have to take 
into account that DRG are typically dusty objects, such that we should 
probably normalise this fading to their unobscured 
luminosity. Assuming that the typical reddening of DRG is 
$E(B-V)\simeq 0.75\pm0.25$ with a Calzetti extinction curve, and that 
they evolve to present-day objects with little dust extinction, we 
find that the typical change in rest-frame magnitude $\Delta M_I= 
M_I(z)-M_I(z=0)$ is $0.26\pm 0.65$ at $<z>=1.5$ and $-0.49\pm 0.65$ at 
$<z>=2.7$. Given the average rest frame luminosities described above, 
this would imply that the descendents of DRG in this simple model have 
rest frame luminosities of about $M_I(z=0)=-22.56$ and $-22.71$.  The 
typical $M^*$ magnitude in the I band for local galaxies in the SDSS 
is $M_i=-22.48$ (\cite{blanton01}), which increases to $-23.2$ if one 
considers only the reddest galaxies ($g-r\ge 0.74$). Considering that 
it is obviously implausible that all DRGs are observed at the end of 
their star--forming phase, and that therefore they will end up in more 
luminous and massive objects than predicted by this exercise, one can 
conclude that both the low-$z$ and high-$z$ DRGs are consistent with 
being the progenitors of local massive galaxies. The analysis of 
clustering will help to clarify this conclusion. 
 
%______________________________________________________________ 
 
\section{Spatial distribution of DRGs: the clustering properties} 
 
It is already known that DRGs are not uniformly distributed on the 
sky, but they are clustered on scales of several Mpc. The analysis of 
the HDFS shows that the DRGs are in prevalence concentrated in one 
quadrant of the WFPC, while in the HDFN there is only one DRG, 
suggesting that this population could be strongly clustered and 
affected by cosmic variance (\cite{vanzhdfs,franx,daddi}), such that  
the small area covered by surveys like HDFN or FIRES  
prevents to derive a robust 
measurement of their clustering properties and their redshift 
evolution.   
 
We therefore present in the following a detailed analysis of the 
clustering properties of our GOODS-MUSIC DRG sample. Thanks to the 
available statistics, we will consider both the overall sample, 
similarly to what already done in previous works, but we will also 
divide the sample into two different sub-groups: the first one, 
containing
objects with $1<z<2$, where the dusty starburst population is expected 
to be the dominant component; the second one, containing objects with 
$2<z<4$, where also relatively evolved galaxies are represented in the sample. 
 
\begin{figure} 
\includegraphics[width=9cm]{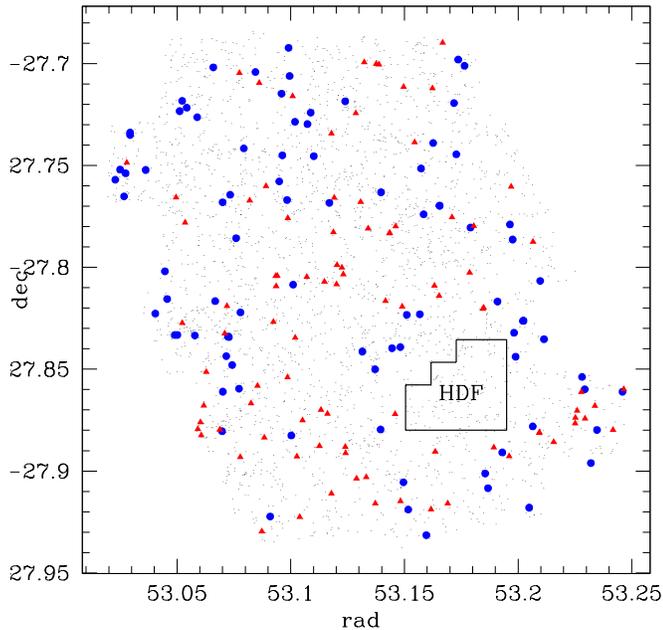} 
\caption{The angular distribution of selected DRGs 
in the GOODS-South Field. The symbols are coded according to the 
redshift: DRGs at $z\ge 2$ and at $z\le 2$ are shown by red triangles 
and blue circles, respectively; black dots refer to normal 
galaxies at all redshifts.  For comparison,  the size of the HDF is 
also shown.  The DRGs are clustered and not uniformly 
distributed over areas larger than the HDFs: this shows that the cosmic 
variance for DRGs is dramatic at small scales.} 
\label{radec} 
\end{figure} 
 
\subsection{Angular Two Point Correlation Function} 
 
We have used the Landy-Szalay estimator (\cite{ls}) to obtain the 
two-point correlation function (TPCF) in the angular coordinates 
($\alpha$,$\delta$) for DRGs in the GOODS Field: 
 
\begin{equation} 
w(\theta)=\frac{GG(\theta)-2GR(\theta)+RR(\theta)}{RR(\theta)} \ , 
\end{equation} 
where $GG(\theta)$ is the number of observed galaxy pairs at distance 
between $\theta$ and $\theta+d\theta$, $GR(\theta)$ is the number of 
observed-random pairs and $RR(\theta)$ is the random-random pairs. We 
compute $GR$ and $RR$ as mean values of 1,000 simulated random 
catalogs.  The random sample of galaxies is obtained by randomly 
generating the coordinates ($\alpha$,$\delta$) in the GOODS-CDFS 
field. Each random galaxy is then retained or rejected according to 
the magnitude limit at the selected position.  This ensures to 
correctly reproduce the selection function of observed DRGs, even in 
presence of a complicated exposure map, like the GOODS survey one 
(\cite{grazian}). Finally we correct the observed $w(\theta)$ taking 
into account the bias arising from the finite boundary of the sample 
(see the details in Appendix A). 
 
Errors on the angular correlation function, $\sigma_w$, are determined 
by Poisson statistics, through the relation 
\begin{equation} 
\sigma_w=\sqrt{\frac{1+w(\theta)}{GG(\theta)}} \ . 
\end{equation} 
 
We fit the angular correlation function (computed in annuli of 
increasing $\theta$) by a power-law relation, 
$w(\theta)=(\theta/\theta_0)^{-\delta}$, fixing $\delta=0.8$. 
Following \cite{croft} (see also \cite{croom,aerqs3}) the fit is 
carried out by using a Maximum Likelihood Estimator (MLE) based on 
Poisson statistics and unbinned data.  A detailed description of the 
MLE can be found in Appendix A. 
 
\begin{figure} 
\includegraphics[width=9cm]{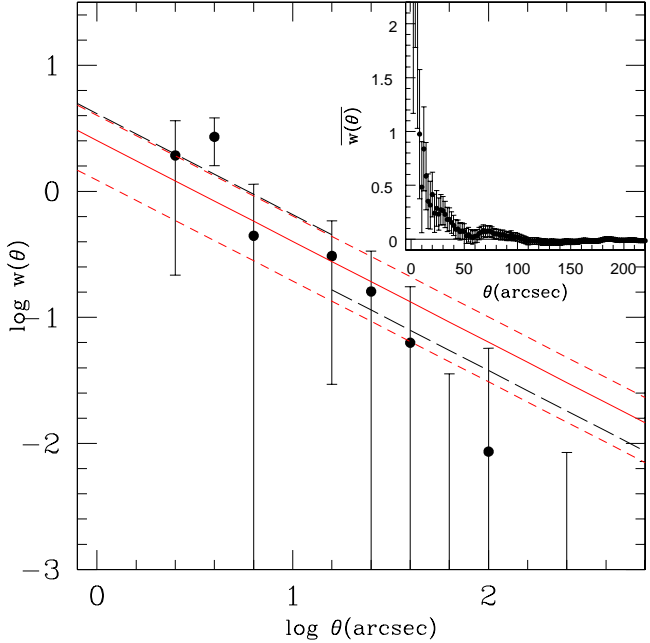} 
\caption{{\em Large panel}: 
the differential angular TPCF (in a logarithmic scale) for the DRGs in 
the GOODS field (filled circles with 1$\sigma$ error bars). We also 
plot the MLE best fit power-law relation (solid line) with its 
1$\sigma$ confidence interval (short-dashed line), as computed in the 
interval $1\le\theta\le 100$ arcsec: the corresponding correlation 
scale of DRGs is $\theta_0=3.19$ arcsec. The two dotted lines refer to 
the MLE best fits on limited intervals: $\theta\le 10$ arcsec and $\theta\ge 
10$ arcsec. Note that at small scales the TPCF is enhanced 
by the presence 
of multiple galaxies in the same DMH, while at large scales the 
boundary effect may become critical. {\em Small quadrant}: the angular TPCF 
integrated over circles of increasing radii (filled circles with 
1$\sigma$ error bars).} 
\label{clust} 
\end{figure}

The results for the DRG TPCF are presented in Fig. \ref{clust} (large 
quadrant), together with the MLE fit with the corresponding 1$\sigma$ 
confidence intervals.  Considering the interval $1\le\theta\le 100$ 
arcsec, we find a clustering scale of $\theta_0=3.19^{+2.48}_{-1.90}$ 
arcsec.  The mean redshift and absolute magnitude for the clustered 
galaxies are $z_{\xi}=2.1$ and $M_I=-22.8$, respectively.  The small 
quadrant of Fig. \ref{clust} shows the TPCF integrated in circles of 
increasing apertures $\theta$.  We do not use this quantity to 
fit the best value for $\theta_0$, since errors are correlated in different 
bins of angular separation. However, we can obtain from its 
value an indication of the clustering strength: at $\theta=12$ arcsec 
we observe 23 pairs, while simulations of random distributions predict 
12 pairs only, which is a detection at about 3$\sigma$; at 
$\theta=6$ arcsec we derive an excess of 7 pairs over 3 random, which 
represents a 4$\sigma$ detection. 
 
By looking at the integrated angular TPCF shown in the small quadrant 
of Fig. \ref{clust}, we notice that it is still significantly 
non null even at large scales ($\theta \approx 50-60$ arcsecs), which 
are comparable to the angular size of the HDFs. This result confirms 
that the difference in the DRG number density found in previous 
surveys is due to both the cosmic variance and their strong 
clustering, whose effects can become dramatic when considering deep 
pencil beam surveys, which are conducted over small areas, like the 
HDFs or the Hubble Ultra Deep Field (HUDF, \cite{udf}). 
 
To have a look at the redshift evolution of the DRG clustering 
properties, we compute the correlation scale of the low- and 
high-redshift samples, separately, and find a clear evidence of a 
strong evolution: we indeed estimate a correlation scale of 
$\theta_0=3.69_{-3.35}^{+5.03}$ arcsec ($z_\xi=1.5$ and $M_I=-22.30$) 
for the low-redshift sample (76 galaxies), and 
$\theta_0=13.68_{-6.29}^{+7.84}$ arcsec ($z_\xi=2.7$ $M_I=-23.20$) 
for the high-redshift one (88 galaxies).

We note that it is well known that at small scales ($\theta\le 
10$ arcsec) the TPCF is dominated by substructures, produced by the 
existence of multiple galaxies inside massive halos (see, 
e.g. \cite{lee}).  This effect is also evident in Fig. \ref{radec}, 
where the presence of close-by galaxy pairs or triplets is clearly 
visible.  To measure the clustering properties of dark matter halos 
(DMHs) hosting DRGs, it is necessary to avoid using only the smallest 
scales, where the halo occupation distribution (HOD) is plausibly 
larger than unity.  Using the total DRG sample, we obtain a 
correlation length of $\theta_0=5.89^{+3.74}_{-3.10}$ arcsec for 
$\theta\le 10$ arcsec, while in the interval $10\le \theta\le 100$ the 
TPCF is significantly weaker, with a MLE fit of 
$\theta_0=1.67^{+2.17}_{-1.50}$ arcsec (see the long dashed lines in 
Fig. \ref{clust}).  It is important to remark, however, that the 
redshift evolution is clearly detected at both scales, although the 
uncertainties become obviously much larger. At the scale of 
$\theta\le 10$, indeed, the correlation length is 
$\theta_0=3.84^{+7.15}_{-3.46}$ arcsec at $1<z<2$ and 
$\theta_0=15.52^{+9.28}_{-7.60}$ arcsec at $2<z<4$.  At $10\le 
\theta\le 100$, the correlation length is 
$\theta_0=2.89^{+3.90}_{-2.65}$ arcsec at $1<z<2$ and 
$\theta_0=8.48^{+13.20}_{-6.72}$ arcsec at $2<z<4$.  Notice that in 
our following discussion we will use the clustering length obtained by 
the fit over the global range $1\le\theta\le 100$ arcsec, since it is 
a robust compromise against boundary effects at the largest scales and 
against HOD effect at the smallest scales. 
 
\subsection{Spatial clustering} 
 
To convert the angular correlation length to physical units we can 
invert the angular TPCF through the Limber equation (\cite{limber}), 
adopting  the DRG 
redshift distribution presented in Fig. \ref{zhist}. 
Leaving the detailed calculations to Appendix A, we have: 
\begin{equation} 
w(\theta)=\frac{r_0^\gamma \theta^{1-\gamma} I(\gamma) 
\int_0^{\infty} (\frac{dN}{dz})^2 r(z)^{1-\gamma}(\frac{dz}{dr}) 
dz}{N^2_{\rm obj}} \ , 
\end{equation} 
where $I(\gamma)=3.67909$ when $\gamma=1.8$ is assumed. 
 
Using the value for $\theta_0=3.19^{+2.48}_{-1.90}$ derived through 
the MLE fit to the angular TPCF, for the complete DRG sample, we 
obtain a correlation length of $r_0=9.78^{+2.85}_{-3.24} h^{-1}$ Mpc. 
Using the same Limber equation, the corresponding comoving correlation 
lengths are $r_0=7.41^{+3.45}_{-4.84} h^{-1}$ Mpc and 
$r_0=13.36^{+2.99}_{-3.20} h^{-1}$ Mpc, for the sub-samples at $1<z<2$ 
and $2<z<4$, respectively. 
 
We note that the TPCF for the higher-redshift sub-sample is different 
with respect to the value obtained by \cite{daddi} for DRGs in the 
HDFS, although still marginally consistent, because of the relatively 
large error budget.  We have to notice, however, that for their 
analysis they applied a colour selection criterion which is bluer 
($J-Ks\ge 0.7$) than the one adopted here ($J-Ks\ge 1.3$).  We tested 
that, by selecting in the GOODS region DRGs at $2\le z\le 4$ with 
their same colour cut, we obtain a sample of 232 galaxies, with a 
typical redshift of $z_{\xi}=2.9$, having a correlation length of 
$r_0=8.8\pm1.7 h^{-1}$ Mpc, which is comparable to the value provided 
by \cite{daddi} ($8.3\pm1.2 h^{-1}$ Mpc). A redder cut ($J-Ks\ge 1.3$) 
applied for DRGs in the HDFS actually results in a larger correlation 
length of $r_0=14.5^{+3.1}_{-3.7} h^{-1}$ Mpc (\cite{daddi}), which is 
consistent with our estimate. 
 
As a further comment, we also notice that the error associated to our 
estimate for $r_0$ in our whole sample ($\sim 3 h^{-1}$ Mpc) is 
slightly higher than the value quoted by \cite{daddi} for the DRGs in 
the HDFS, even if the samples have a different number of objects (197 
DRGs in GOODS against 49 in the HDFS).  This is due to the fact that 
we include in the error budget the effects of cosmic variance, which 
is the dominant effect in this kind of study and it is not included in 
the error bars quoted for DRGs in HDFS. 
 
It is interesting to compare these results with other estimates of 
clustering strength, for other related classes of objects.  In order 
to avoid the dependence of the scale length $r_0$ on the power-law fit 
$\gamma$, it can be useful to present the results in a non-parametric 
form. This can be done by using the quantity $\bar{\xi}$, defined as 
the correlation function $\xi(r)=(r/r_0)^{-\gamma}$ integrated over 
a sphere of a given radius $r_{\rm max}$: 
\begin{equation} 
\bar{\xi}(r_{\rm max})=\frac{3}{r_{\rm max}^3}\int_0^{r_{\rm max}} 
\xi(x)x^2{\rm d}x \ . 
\end{equation} 
In general, the larger the scale on which the clustering is measured, 
the easier the comparison with the linear theory of the structure 
evolution. Since in the following we want to compare our results with 
those obtained for different values for $\gamma$, we prefer to quote 
clustering amplitudes within $20 h^{-1}$ Mpc, a scale for which 
linearity is expected to better than a few per cent.  Choosing a large 
radius also reduces the effects of small scale peculiar velocities and 
redshift measurement errors, which can be a function of redshift.  
 
\begin{figure} 
\includegraphics[width=9cm]{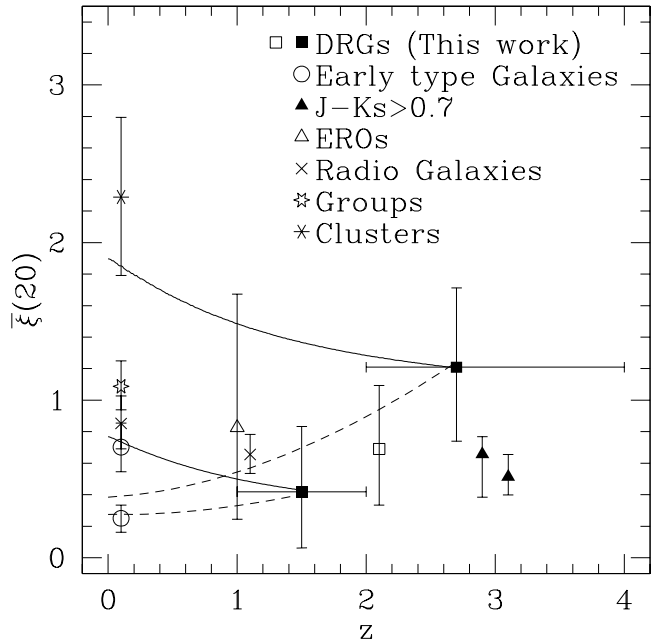} 
\caption{ 
The integrated clustering strength $\bar\xi(20h^{-1}Mpc)$ as a 
function of redshifts for different objects: DRGs, EROs, powerful 
radio-galaxies, ellipticals and galaxy groups/clusters. 
Filled squares show the results  for low- and high-$z$ DRGs in the GOODS 
region, while void square represents the whole DRG sample. 
The solid lines show the 
predicted evolution of the clustering according to the {\em 
object-conserving} model, tuned to the DRGs at low- and high-$z$, while 
the dashed lines reproduce the clustering evolution according to the 
{\em merging model}. The plot suggests that high-redshift  
DRGs can be the progenitors of local ellipticals, but may 
evolve into more massive objects, like EROs at $z\sim 1$ and 
groups/clusters of galaxies in the local universe. The horizontal 
error bars show the redshift intervals for the DRGs in this work. 
Filled triangles 
show the values of the correlation strength for DRGs with $J-Ks\ge 
0.7(AB)$ as estimated in the HDFS (\cite{daddi}) at $z=3.1$ and in the 
GOODS region at $z=2.9$ (this work).} 
\label{xiz} 
\end{figure}

Fig. \ref{xiz} compares the values for $\bar\xi(20h^{-1})$ that we 
obtained for DRGs in the GOODS field (summarised in Table 
\ref{clusttab}) to the corresponding estimates for other classes of 
objects, both at low and high redshift. It is immediately clear that 
the high-redshift ($z>2$) sample of DRG is drawn from a remarkably 
highly clustered population, most likely more clustered than the $z<2$ 
DRG population. 
 
At $z=0$, the only galaxies having correlation lengths as large as 
10-11 $h^{-1}$ Mpc (corresponding to $\bar\xi(20h^{-1})\sim 1$) are 
morphologically-selected giant ellipticals or radio-galaxies. 
\cite{guzzo} estimate $r_0=8.35\pm0.76h^{-1}$ Mpc for 
early-type galaxies with $M_B\le -19.5+5log(h)$ in the Pisces-Perseus 
super-cluster survey, while \cite{adami} derive a significantly smaller 
value ($r_0=7 h^{-1}$ Mpc with $\gamma =-1.79$) from the SSRS2 
redshift survey.  The discrepancy between these two measurements is 
probably originated by the presence in the first survey of the 
super-cluster, which enhances the correlation function. 
\cite{overzier} and \cite{rottgering} find that 
local radio-galaxies have large clustering lengths (see also 
\cite{peakcock}) and that the high degree of correlation between 
hosting ellipticals and luminous radio-sources suggests an interesting 
possible comparison for distant samples.   
 
For small groups of galaxies in the local Universe, the typical value for  
$\bar\xi$ has been measured by \cite{girardi,zandivarez,padilla}, and 
again shown in Fig.\ref{xiz}. 
\cite{collins} report the results of the spatial two-point correlation 
function for the galaxy cluster survey ROSAT-ESO Flux-Limited X-ray 
(REFLEX), finding $\bar\xi(20h^{-1})=2.29\pm0.50$ for rich clusters 
at $z\le 0.3$. 
 
More ordinary elliptical galaxies show a range of clustering strength, 
that is strongly dependent on the absolute magnitude. We reproduce in 
Fig. \ref{xiz} the range corresponding to local elliptical galaxies 
ranging from $M_B=-17$ to $M_B=-21$, taken from \cite{2dfgrs} and 
\cite{sdss}. 
 
At high redshifts, we also display the values observed for EROs 
(\cite{daddieros}) and for bluer DRG (\cite{daddi}). 
 
This compilation of clustering strength for a wide range of objects 
shows that DRG are among the mostly clustered objects at galactic 
scales, and suggests that they might be related to the progenitors of 
similarly clustered objects at lower redshifts, as EROs or local 
massive ellipticals.  Unfortunately, a firm conclusion in this context 
is not straightforward, since we do not know the evolution of the bias 
parameter for this class of high-redshift objects. This point will 
be better discussed in the final section. 
 
\begin{table} 
\caption[]{Clustering properties of DRGs} 
\begin{tabular}{lccccc} 
\hline 
\hline 
$Type$ & $r_0$ & $\gamma$ & $\bar\xi(20h^{-1})$ & $z$ & $M_I$ \\ 
       & ($h^{-1}$ Mpc) &          &           &  \\ 
\hline 
DRGs & $9.78^{+2.85}_{-3.24}$ & 1.8 & $0.690^{+0.403}_{-0.356}$ 
& 2.1 & -22.8 \\ 
low-$z$ DRGs & $7.41^{+3.45}_{-4.84}$ & 1.8 & $0.419^{+0.414}_{-0.357}$ 
& 1.5 & -22.3 \\ 
high-$z$ DRGs & $13.4^{+2.99}_{-3.20}$ & 1.8 & $1.209^{+0.530}_{-0.470}$ 
& 2.7 & -23.2 \\ 
$J-Ks\ge 0.7$ & $8.77^{+1.62}_{-1.70}$ & 1.8 & $0.657^{+0.112}_{-0.272}$ 
& 2.9 & -23.0 \\ 
\hline 
\hline 
\end{tabular} 
\label{clusttab} 
\end{table} 
 
%______________________________________________________________ 
 
\section{Summary and Discussion} 
  In this paper we have presented an analysis of Distant Red
Galaxies  (DRG) selected in the GOODS-South region. In particular, we
have used  the GOODS-MUSIC sample, that has been compiled from a
unique dataset  that comprises accurate multi-wavelength coverage (14
bands from $0.3$  to $8 \mu m$) of $\sim$3000 galaxies in $Ks$
complete sample, with  accurate estimates of the photometric
redshifts for {\em all} galaxies  in the field. From the GOODS-MUSIC
sample, we have selected 179 DRGs according to the criterion
proposed by \cite{franx}, $J-Ks\ge 1.3$ at a typical magnitude
limit of $Ks=23.5(AB)$ and down to $Ks=23.8$ in a limited area.
The wide and deep covered area (135 
sq. arcmin), together with the extended SED information and the 
precision in photometric redshifts ($\sigma_z=0.06$), allows to
study  the statistical properties of DRGs, like the redshift
distribution,  number density and clustering properties at an
unprecedented level. 

The derived number density is consistent with that found by 
\cite{labbe03}, with approximatively 1 DRG per sq. arcmin. at 
$Ks=23.5$. The redshift distribution shows a smoothed peak around 
$z\sim 2$, with extended tails both to $z=1$ and $z=4$. Bright DRGs 
($Ks\le 22$) tend to dominate the $z\sim 1$ region, while apparently 
faint DRGs ($Ks > 22$) are distributed widely around $z\sim 2.0-3.5$. 
The two populations also have different intrinsic properties: 
low-redshift DRG are slightly less luminous than their higher-$z$ 
counterparts ($<M_I>=-22.3$ and $<M_I>=-23.2$, respectively), and 
possibly slightly less massive ($<M_{star}>=8.15\cdot 10^{10} M_\odot$ 
and $<M_{star}>=9.90\cdot 10^{10} M_\odot$, respectively). 
 
In particular, we investigated on the spatial distribution of DRGs 
through the Two-Point Correlation Function (TPCF) analysis. We find 
that DRGs from the overall sample are significantly clustered 
(4$\sigma$ detection), with a typical correlation length of 
$\theta_0=3.19^{+2.48}_{-1.90}$ arcsec, corresponding, through the 
Limber equation and the observed redshift distribution, to 
$r_0=9.78^{+2.85}_{-3.24}h^{-1}$ Mpc.  We also find that the 
clustering strength of DRGs increases with the $J-Ks$ colour cut used 
for selection. 
 
Using the relatively large sample of DRG provided by the GOODS-MUSIC 
sample, we divided the DRG sample in two sub-groups in redshift, one 
with $1\le z\le 2$ and the other with $2\le z\le 4$. The clustering of 
low-$z$ DRGs is significantly lower than that of the high-$z$ DRGs, 
with $r_0=7.41^{+3.45}_{-4.84}h^{-1}$ Mpc and 
$13.4^{+2.99}_{-3.20}h^{-1}$ Mpc, respectively. 
It is useful to stress here that this behaviour is not due to a physical 
evolution of the DRG population. It is the result of a
selection criterion which provides an heterogeneous group of dusty starburst 
and massive/evolved galaxies with different redshift distribution. 
 
Unfortunately, a direct comparison of the clustering properties of
DRGs with those of other objects can be misleading, since it is not
known a priori the connection between these classes. However, it is
possible to constrain the clustering evolution of the descendents of
the DRG population using two extreme, simplified models, as proposed
by \cite{matarrese} and \cite{moscardini} for the merging of
galaxies in a $\Lambda$CDM hierarchical clustering scenario.
In one case, that was named
{\em object-conserving model}, we assume that the observed DRGs do not
undergo any subsequent phase of merging with other objects, included
those of lower mass. This model, which is conceptually close to a sort
of ``passive evolution'' scenario, assumes that the
galaxies form at some characteristic redshift by some non-linear
process which induces a bias parameter at that epoch, and that their
subsequent motion is purely caused by gravity, following the
continuity equation. An obvious consequence of this model is that the
bias factor will not be constant for all time, but will tend to unity
as time goes on because the galaxies will be dragged around by the
surrounding density fluctuations, populated by less clustered objects.
This scenario, which corresponds to have an extremely long merging or
disruption time, provides an upper limit to the evolution of the
clustering properties of DRG descendents, and is shown as thick solid
lines in Fig.\ref{xiz}, after normalisations to the DRG values
obtained in this paper. On the other side, we use a {\em merging
model}, where the - even more extreme - assumption is that galaxies
continue the merging process down the lowest redshifts, with the same
(high) merger rate of their parent halos. This clearly extreme model
provides a lower limit of the evolution of the clustering properties
of DRG descendents, and is shown as dashed lines in
Fig.\ref{xiz}. These theoretical predictions have been obtained
adopting the standard $\Lambda CDM$ power spectrum, normalised to
reproduce the local cluster abundance ($\sigma_8=0.9$).
 
Although the error budget on the estimate of $\bar\xi(20h^{-1})$ on 
the two DRG samples is still relatively large, we can use these two 
limiting theoretical predictions to attempt a physical interpretation 
of our results. 
 
First, the observed value of $\bar\xi(20h^{-1})$ for the low-$z$ DRGs 
is outside the range predicted for the evolution of the higher $z$ 
sample: this suggests that is unlikely that the two samples are drawn 
from the same population, observed at two different stages of 
evolution.  
 
If we look at the low redshift range predicted for the DRG evolution, 
it is suggested that high-redshift DRGs (i.e. those typically selected 
at $Ks>22$, see Fig.\ref{zhist}) likely represent the progenitors of 
the more massive galaxies in the local Universe, i.e. the more 
luminous ellipticals, and might mark the regions that will later 
evolve into structures of intermediate mass, like groups or small 
clusters. 
 
On the other hand, low-redshift DRGs (i.e. those typically selected 
at $Ks<22$), will likely evolve into slightly less massive field galaxies, 
approximately around the characteristic luminosity $L^*$ of local 
ellipticals. 
 
Our observations provide further evidence for the so called 
``downsizing'' scenario that has emerged in many different aspects of 
high redshift galaxies, providing evidences that more massive galaxies 
have formed preferentially at higher redshifts than less massive 
ones. Here we find the same trend, since high redshift DRGs are more 
clustered, more luminous, and most likely to evolve into more massive 
galaxies than their lower-$z$ counterparts.
 
%______________________________________________________________ 
 
\begin{acknowledgements} 
It is a pleasure to thank the GOODS Team for providing all the
imaging material 
available worldwide. Observations were carried out using the Very Large 
Telescope at the ESO Paranal Observatory under Program IDs LP168.A-0485 and 
ID 170.A-0788. We are grateful to the referee for useful and constructive 
comments. 
\end{acknowledgements}

\appendix 
 
\section{The Two-Point Correlation Function and the Limber Equation} 
 
The calculation of the TPCF over small regions of the sky is affected 
by boundary effects. This bias, known as the integral constraint, is 
produced by the fact that the angular TPCF is computed over a limited 
area $\Omega$: the consequence is a reduction of the amplitude of the 
correlation function by  
 
\begin{equation} 
w_{\Omega}=\frac{1}{\Omega^2}\int \int w(\theta)d\Omega_1 d\Omega_2 \ . 
\end{equation} 
 
Following \cite{roche02} we estimated $w_{\Omega}$ numerically as 
\begin{equation} 
w_{\Omega}=k A_w= 
\frac{\sum{RR(\theta)A_w\theta^{-\delta}}}{\sum{RR(\theta)}} \ , 
\end{equation} 
where we assumed for $w(\theta)$ a power-law relation: $w(\theta)=A_w 
\theta^{-\delta}$. Fixing $\delta=0.8$ we obtain $w_{\Omega}=10.692 
A_w$ and the corrected expression for the angular TPCF becomes 
\begin{equation} 
w(\theta)=w^{\rm obs}(\theta)+w_{\Omega} \ . 
\end{equation} 
 
The fit of the differential angular TCPF (corrected for boundary 
effects and computed in annuli of increasing $\theta$) is carried out 
by using a Maximum Likelihood Estimator (MLE), described in 
\cite{croft}. This method is based on Poisson statistics and unbinned 
data.  Unlike the usual $\chi^2$ minimisation, MLE avoids the 
uncertainties related to the bin size, the position of the bin centre 
and the bin scale (linear or logarithmic). 
 
To build this estimator, it is necessary to estimate the predicted 
probability distribution of galaxy pairs, given a choice for the 
correlation length $\theta_0$ and the slope $\delta$.  By using all 
the distances between the random-random pairs $RR(\theta)$, we can 
compute the number of pairs $g(\theta)d\theta$ in arbitrarily small 
bins $d\theta$ and use it to predict the expected mean number of 
galaxy-galaxy pairs $h(\theta)d\theta$ in that interval as 
\begin{equation} 
h(\theta)d\theta=[1+w(\theta)] g(\theta)d\theta\ , 
\end{equation} 
where the correlation function $w$ is modelled by assuming a power-law 
expression, $w(\theta)=(\theta/\theta_0)^{-\delta}$, $\delta=0.8$.  In 
this way, it is possible to use all the distances between the $N_p$ 
galaxy-galaxy pairs data to build a likelihood.  In particular, the 
likelihood function ${\cal L}$ is defined as the product of the 
probabilities of having exactly one pair at each of the intervals 
$d\theta$ occupied by the galaxy-galaxy pair data and the probability 
of having no pairs in all remaining intervals. Assuming a Poisson 
distribution, one finds 
\begin{equation} 
{\cal L}= \prod_i^{N_p} \exp[-h(\theta)d\theta] h(\theta)d\theta
\prod_{j\ne i} \exp[-h(\theta)d\theta]\ , 
\end{equation} 
where the index $j$ runs over all the intervals $d\theta$ where there 
are no pairs.  As usual, it is convenient to define the quantity 
$S\equiv -2 \ln {\cal L}$, which can be re-written, once we retain 
only the terms explicitly depending on the unique model parameter 
$\theta_0$, as 
\begin{equation} 
S=2\int^{\theta_{\rm max}}_{\theta_{\rm min}} h(\theta)d\theta -2\sum_i^{N_p} 
\ln[h(\theta_i)]\ . 
\end{equation} 
The integral in the previous equation is computed over the range of 
scales where the fit is made. The minimum scale is set by the smallest 
scale at which we find a DRG pair (in our case $\theta_{\rm min}=0.6$ 
arcsec), while for the maximum scale we adopt $\theta_{\rm max}=15$ 
arcsec.  The latter choice is made to avoid possible biases from large 
angular scales, where the signal is weak.  By minimising $S$ it is 
possible to obtain the best-fitting parameter $\theta_0$. The 
confidence level is defined by computing the increase $\Delta S$ with 
respect to the minimum value of $S$. In particular, assuming that 
$\Delta S$ is distributed as a $\chi^2$ with one degree of freedom, 
$\Delta S=1$ corresponds to 68.3 per cent confidence level. It should 
be noted that by assuming a Poisson distribution the method considers 
all pairs as independent, neglecting their clustering. Consequently 
the resulting error bars can be underestimated (see the discussion in 
\cite{croft}). 
 
To convert the TPCF from angular to spatial (3D) coordinates we can 
resort to the so-called Limber equation. Its original formulation is 
given by: 
\begin{equation} 
w(\theta)=\frac{\int_0^{\infty} \Psi(r_1) r_1^2 dr_1 \int_0^{\infty} \Psi(r_2) 
r_2^2 \xi(r_{12}) dr_2}{N_{\rm obj}^2} \ , 
\label{eq:limber} 
\end{equation} 
where 
\begin{equation} 
r_{12}^2=[r_1^2+r_2^2-2 r_1 r_2 \cos(\theta)] \ . 
\end{equation} 
 
Adopting the new variables 
\begin{equation} 
r=\frac{r_1+r_2}{2}; y=\frac{r_1-r_2}{r\theta} \ , 
\end{equation} 
and, assuming the small angle approximation, 
we obtain,  
\begin{equation} 
r_{12}=r\theta(1+y^2)^{1/2} \ , 
\end{equation} 
which, when substituted in Eq.(\ref{eq:limber}), gives 
\begin{equation} 
w(\theta)=\frac{\theta \int_0^{\infty} \Psi^2(r) r^5 dr 
\int_{-\infty}^{+\infty} \xi[r\theta(1+y^2)^{1/2}] dy}{N_{\rm obj}^2} \ . 
\end{equation} 
Finally, using the symmetric properties of $\xi(r)$, the expression 
for the Limber equation becomes: 
\begin{equation} 
w(\theta)=\frac{\theta^{1-\gamma} I(\gamma) \int_0^{\infty} \Psi^2(r) 
r^5 (\frac{r}{r_0})^{-\gamma} dr}{N_{\rm obj}^2} \ , 
\end{equation} 
where 
\begin{equation} 
I(\gamma)\equiv \sqrt{\pi} 
\frac{\Gamma(\frac{\gamma-1}{2})}{\Gamma(\frac{\gamma}{2})}=3.67909 \ , 
\end{equation} 
when the usual value $\gamma=1.8$ is adopted. 
 
The redshift distribution of real objects can be written as 
\begin{equation} 
\frac{dN}{dz}= \Psi(r) r^2 \frac{dr}{dz} \ , 
\end{equation} 
where the variation of redshift with comoving distance for a 
$\Lambda CDM$ model is given by 
\begin{equation} 
\frac{dz}{dr}= \frac{H_0\sqrt{\Omega_M(1+z)^3+\Omega_\Lambda}}{c} \ . 
\end{equation} 
 
If for the angular correlation function we assume a power-law relation 
$w(\theta)=A\theta^{-\delta}$, with $\delta=\gamma-1$,  
it is easy to invert the Limber equation, 
through the assumption of a constant value for $r_0$ with redshift: 
\begin{equation} 
w(\theta)=\frac{\theta^{1-\gamma} I(\gamma) r_0^{\gamma} 
\int_0^{\infty} (\frac{dN}{dz})^2 [r(z)]^{1-\gamma} (\frac{dz}{dr}) 
dz}{N_{\rm obj}^2} \ . 
\end{equation} 
If $\xi(r)=(\frac{r}{r_0})^{-\gamma}$ is the TPCF in 3D space, we 
obtain $w(\theta)=(\frac{\theta}{\theta_0})^{1-\gamma}$ for the TPCF 
in the angular coordinates. 
 
\end{document}